# Macro-molecular data storage with petabyte/cm$^3$ density, highly parallel read/write operations, and genuine 3D storage capability

Masud Mansuripur and Pramod Khulbe
Optical Sciences Center, University of Arizona, Tucson, Arizona 85721, masud@u.arizona.edu

**Abstract**. Digital information can be encoded in the building-block sequence of macro-molecules, such as RNA and single-stranded DNA. Methods of "writing" and "reading" macromolecular strands are currently available, but they are slow and expensive. In an ideal molecular data storage system, routine operations such as write, read, erase, store, and transfer must be done reliably and at high speed within an integrated chip. As a first step toward demonstrating the feasibility of this concept, we report preliminary results of DNA readout experiments conducted in miniaturized chambers that are scalable to even smaller dimensions. We show that translocation of a single-stranded DNA molecule (consisting of 50 adenosine bases followed by 100 cytosine bases) through an ion-channel yields a characteristic signal that is attributable to the 2-segment structure of the molecule. We also examine the dependence of the rate and speed of molecular translocation on the adjustable parameters of the experiment.

**1. Introduction**. Many of the traditional problems in disk and tape data storage can be overcome if data-blocks were to be released from the confines of a disk (or tape) and allowed to float freely between read/write stations (i.e., heads) and permanent "parking spots." The heads and parking spots thus become fixed structures within an integrated chip, while the macro-molecular data blocks themselves become the (mobile) storage media. In this scheme, a large number of read/write heads could operate in parallel, the heads and parking spots would be constructed (layer upon layer) in a truly 3-dimensional fashion, and individual nanometer-sized molecules – strung together in a flexible macromolecular chain – would be used to represent the 0's and 1's of binary information. In this paper we discuss the potential advantages of this alternative scheme for (secondary) data storage and, to demonstrate the feasibility of the concept, we present results of experiments based on DNA molecules that travel within micro-fluidic chambers.

In principle, the four nucleotides of DNA can be used to represent a 2-bit sequence (A = 00, C = 01, G = 10, T = 11), although practical considerations may impose certain restrictions on the specific sequences that can be used to encode the information. A data storage device built around this concept must have the ability to (i) create macromolecules with any desired sequence of building blocks, i.e., *write* or encode the digital information into macromolecular strands; (ii) analyze and decode the sequence of a previously created macromolecule, i.e., *read* the recorded information; (iii) provide an automated and reliable mechanism for transferring the macromolecules between the read station, the write station, and designated locations (parking spots) for storing each such macro-molecule. Although methods of writing and reading macromolecular strands are currently available (e.g., arbitrary sequences of oligonucleotides can be synthesized, and DNA sequences can be deciphered), these methods require large machines and are slow and expensive. In an ideal molecular data storage system, routine operations such as write, read, erase, store, and transfer must be carried out within an integrated chip, reliably and at high speed.

**2. Device architecture**. As a possible alternative to present-day mass data storage devices (e.g., magnetic and optical disks and tapes), we envision a system in which data blocks are encoded into macromolecules constructed from two or more distinct bases, say, *x* and *y*; the bases can be strung together in arbitrary order such as *xxyxyyxy…xyx* to represent binary sequences of user-data ($x = 0, y = 1$).[1,2] The macromolecular data blocks must be created in a *write station*, transferred to *parking spots* for temporary storage, and brought to a *read station* for decoding and readout. The *erase* operation is as simple as discarding a data block and allocating its parking spot to another macromolecule. (In principle, discarded molecules can be recycled after being broken down to their constituent elements.) The parking spots and read/write stations depicted in Fig. 1, for example, are μ-fluidic chambers connected via μ-channels and μ-valves (not shown) that enable automatic access through an electronic addressing scheme.[2] With the dimensions of the various chambers indicated in Fig. 1, one can readily incorporate, on a $1.0\,cm^2$ surface area, a total of $10^6$ parking spots ($\sim 0.25\,cm^2$), 1000 read/write stations ($\sim 0.1\,cm^2$), and necessary plumbing (e.g., 1μm-wide connecting routes, $1 \times 1\,\mu m^2$ binary valves or switches), which would occupy an area $\sim 0.65\,cm^2$. Assuming megabyte-long data-blocks, the storage capacity of the $10^6$ parking spots in this scheme will be $10^{12}$ bytes/$cm^2$. In a three-dimensional design based on 10μm-thick layers, the capacity of the proposed molecular storage system would exceed $10^{15}$ bytes/$cm^3$. For a comparison with a current state-of-the-art technology, note that storing $10^{15}$ bytes of data on Digital Versatile Disks requires a 128 meter-tall stack of 12cm-diameter DVD platters.

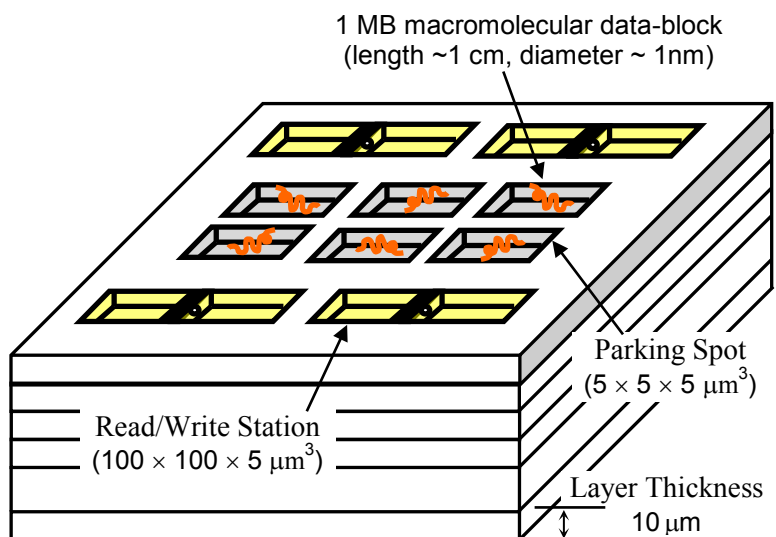

**Fig.1**. Chip surface utilization. The same arrangement of read, write stations and parking spots is repeated in stacked layers.

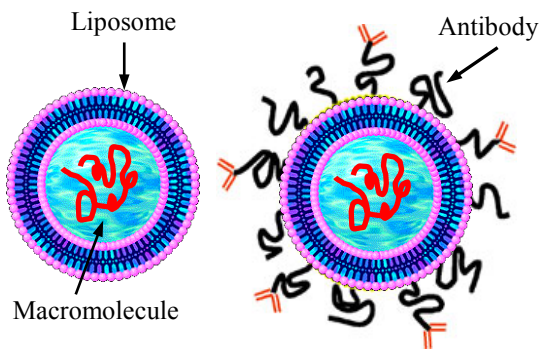

**Fig.2**. Liposomes are microscopic spherical vesicles that form when phospholipids are hydrated. When mixed in water under low shear conditions, the phospholipids arrange themselves in sheets, the molecules aligning side by side with heads up and tails down. The sheets then join tails-to-tails to form a bilayer membrane. Phospholipid bilayer spheres enclosing water and one or more macromolecules can be produced with uniform size (diameter ~ 200 nm or smaller). Antibodies inserted in the outer surface of the liposome can target specific locations within the device tagged with antibody-specific antigens.

One may also envision packaging the molecular data blocks in microscopic spherical vesicles known as liposomes (see Fig. 2). This would provide a robust environment for the transfer of data blocks between the parking spots and the read/write stations. Moreover, by planting specific antibody molecules on the vesicle surfaces, it becomes possible to direct these "addressed" packages to desired locations tagged with antibody-specific antigens (i.e., receptors).

In an effort to demonstrate the feasibility of the proposed molecular storage scheme, we have built miniaturized read stations for DNA strands.[2,3] The idea is to translocate single-stranded DNA molecules through a nano-pore while monitoring the variations of the electrolytic current through the pore, as has been described elsewhere in the context of gene analysis and sequencing.[4-11] Whereas in decoding a genome the DNA sequence could be arbitrary, in the proposed data storage application it is possible to tailor the sequence to the characteristics of the read station. For instance, if the electrical signals corresponding to individual bases turn out to be indistinguishable (due to insufficient signal-to-noise ratio, SNR, or because the ion-channel is too long compared to the dimensions of single nucleotides), one might instead associate the information bits with strings of identical bases; for example, a string of 20 adenosines can be used to represent a 0, while a string of 30 cytosines could represent a 1.

**3. Conventional modes of mass storage and their limitations**. The hallmarks of traditional secondary (i.e., disk, tape) data storage may be summarized as follows: (i) mechanical motion of the storage medium; (ii) mechanical motion of the read/write head; (iii) confinement of data to two-dimensional surfaces; (iv) data organization in sectors or blocks (as opposed to words); (v) difficulty of achieving parallel access to multiple sectors; (vi) dependence of achievable data density on the read/write/servo mechanism (as opposed to the physical properties of the storage medium). By releasing the data-blocks from the confines of the storage medium and giving them an independent physical identity, as depicted in Fig. 3, one could solve many of the traditional problems associated with disk/tape data storage. The released data sectors must still represent large blocks of data (item iv above); they also continue to require mechanical motion for moving around the storage device (item i above). However, the free-floating data blocks can now operate with *stationary* read/write heads, and will no longer be confined to a two-dimensional surface. The read/write heads will be permanently fixed within the storage system (i.e., no need for flying heads, track-following, auto-focusing, or scanning the storage medium to access various sectors). Therefore, the number of read/write stations that can be integrated into the storage system can increase dramatically compared to what is achievable in present-day devices. Parallel access to data thus becomes a natural mode of operation for a system that has hundreds or thousands of independent read/write stations.

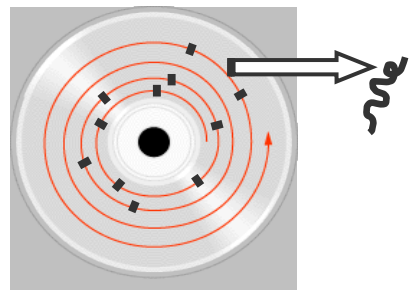

**Fig. 3**. Releasing the data blocks from the storage medium can, in principle, solve many of the problems associated with conventional data storage schemes.

**4. Miniature read station**. In previous publications we have described a method of reading molecular data blocks by translocation through nano-pores.[2] We have also reported methods of fabrication of the requisite fluidic chambers.[2,3] Our current research is focused on miniaturizing the read/write stations and integrating them with parking spots, μ-fluidic channels, and μ-valves. Our read stations, complete with electrodes and access ports, were molded from polydimethylsiloxane (PDMS), cast in a machined Teflon block. This design, which is scalable to even smaller dimensions, allows easy integration with multiple parking spots on the same chip. In the read station depicted in Fig. 4, the needle-like end of the tube that connects the *cis* and *trans* chambers was covered with a Teflon cap (aperture diameter ~20μm) to promote the formation of a lipid membrane over the aperture. The nano-pore, self-assembled in the lipid membrane from seven subunits of α-haemolysin (αHL) protein, is a 10 nm-long ion-channel, whose cap (length ~5 nm) resides in the cis chamber, while its 5nm-long stem spans the lipid membrane. The ion-channel has an opening diameter of 2.6 nm at the entrance to the cap, 1.5 nm in the constricted region in the middle, and 2.2 nm at the far end of the stem located in the trans chamber.[4,5]

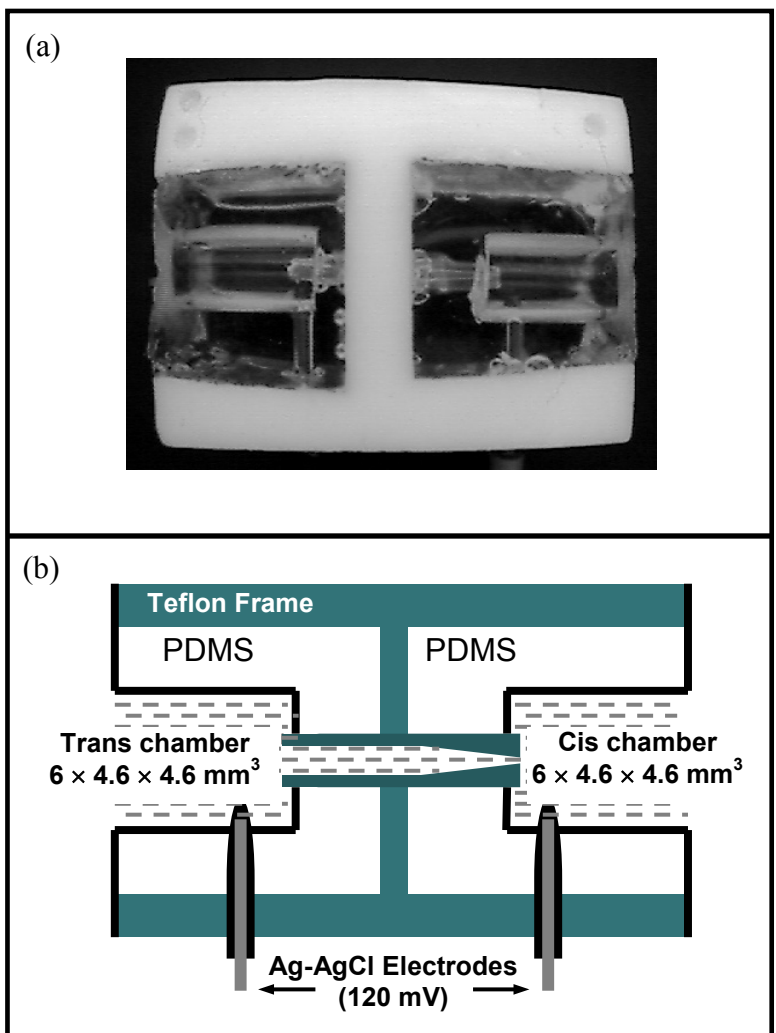

**Fig. 4**. (a) Photograph and (b) diagram of a read station built inside a 10mm-thick H-shaped Teflon frame. A Teflon tube, shrunk at one end to a 20μm diameter aperture, goes through the central wall and forms a tight seal between the *cis* and *trans* chambers. The 6mm-long, 4.6mm diameter cylindrical chambers are large enough to hold the buffer solution for several hours; evaporation reduces the buffer by 50% in 24 hours. The chambers are connected to an Axopatch 200B amplifier via Ag-AgCl electrodes. The 20μm diameter aperture holds a (vertical) lipid bilayer, into which one or more αHL ion-channels are implanted. Single-stranded DNA molecules are driven through the channel by an applied voltage in the range 90-210mV, positive at the trans side.

In the read station of Fig. 4, the cis chamber is accessed (under microscope control) with a micro-pipette from the right-hand side, and the trans chamber from the left-hand side. The micro-pipette is used for filling the chambers with buffer, adding αHL or DNA to the cis chamber, and removing the additives by perfusion. At 50μL volume (partial filling), the chambers are small enough to yield a low-noise electrolytic current signal, yet large enough to avoid problems associated with the evaporation of the liquids. If the device dimensions were to shrink further, both chambers would have to be capped to prevent evaporation.

Our goal was to improve the signal-to-noise ratio (SNR) of readout by increasing *both* the KCl concentration in the buffer and the applied voltage. The increased salt concentration, however, resulted in ion-channel gating, which is unacceptable behavior for observation of DNA translocation events. The higher voltage, on the other hand, helped raise the SNR by enhancing the signal without causing much increase in the noise.

**5. Gating behavior of α-haemolysin nano-pore at high KCl concentration**. The black (top), green (bottom), and red (middle) traces in Fig. 5 show the electrical current through a single nano-pore with 1 molar KCl concentration at applied voltage levels of +210mV, −210mV, and 0 mV, respectively. The channel is always open at this KCl concentration, and the forward current is just over 250pA, while the reverse current is about –200pA. The width of the trace in each case is a good measure of the noise level present during the measurement. (The KCl concentration was 1M on both *cis* and *trans* sides; in other words, there was no transmembrane gradient.)

When the KCl concentration is raised to 2M, the ion-channel exhibits a "gating" behavior, namely, it opens and closes randomly, as can be seen in the blue trace in Fig. 5. The applied voltage was reduced in this case to +120mV to make the forward current (in the open state of the ion-channel) nearly equal to the forward current in the case of 1M KCl concentration. (At 2M KCl, gating behavior was also observed at higher voltages, up to 200mV, and also when a reverse voltage, $V = -120$mV, was applied.) Apparently, at some KCl concentration between 1M and 2M, the nano-pore exhibits gating, irrespective of the level or polarity of the applied voltage.

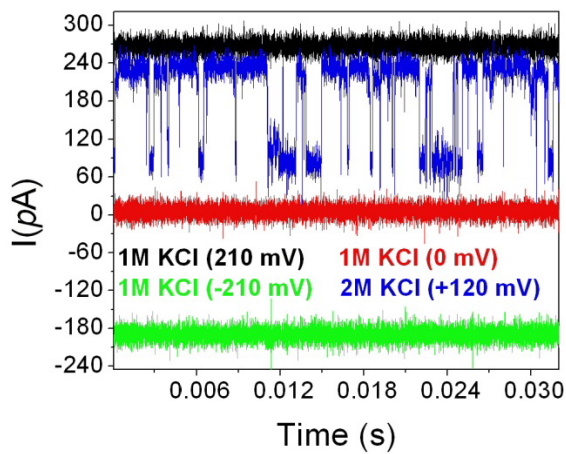

**Fig. 5**. Current traces obtained from an αHL channel incorporated into a bilayer. The red (middle) trace shows the channel in the non-conducting state, when there is no potential difference across the bilayer. The black (top) and green (bottom) traces show the channel in the conducting (open) state in response to a trans-bilayer applied voltage of +210 mV and –210 mV, respectively (in the presence of 1M KCl on both sides of the bilayer). Note the difference in the current level between the black and green traces, which indicates the channel's asymmetric response to a polarity reversal of the applied voltage. In the presence of 2M KCl, and in response to an applied voltage of +120 mV (blue, oscillating trace), the channel fluctuates spontaneously between open and closed states (gating).

**6. Translocation of 5'$A_{50}C_{100}$3' single-stranded DNA**. The ssDNA molecules traversed the ion-channel under $V = 210$mV at 1M KCl buffer concentration, and were observed with the amplifier bandwidth set to 100 kHz. In one trial, of the 171 translocation events monitored, 49 events, or nearly 29%, showed bi-level behavior. (Not every event is expected to exhibit a bi-level signal, either due to the large fluctuations of the blockade current, or because of incomplete translocation, in which the molecule is trapped in the vestibular region of the αHL cap for a certain length of time, then returned to the *cis* chamber without traversing the ion-channel.) The bi-level behavior is seen in three of the four events in Fig. 6. Translocation duration for the entire molecule is typically ~150μs. In some instances, such as the first event in Fig. 6, the molecule appears to get stuck in the midst of translocation, which results in a longer-than-average transition time. The fourth event in Fig. 6 is typical of the remaining 71% of events in which either the molecule entered the pore but did not complete translocation, or the bi-level signal was not clearly visible due to a lack of sufficient SNR. It is worth mentioning that, in our experiments, bi-level signals were *not* observed for $V$ = 120mV, 150mV, or 180mV; the higher applied voltage of 210mV, was somehow necessary to obtain these signals. (Although a bi-level translocation signal from 2-segment RNA molecules has been reported in the literature,[6] we are not aware of any such results in the published record for ssDNA.)

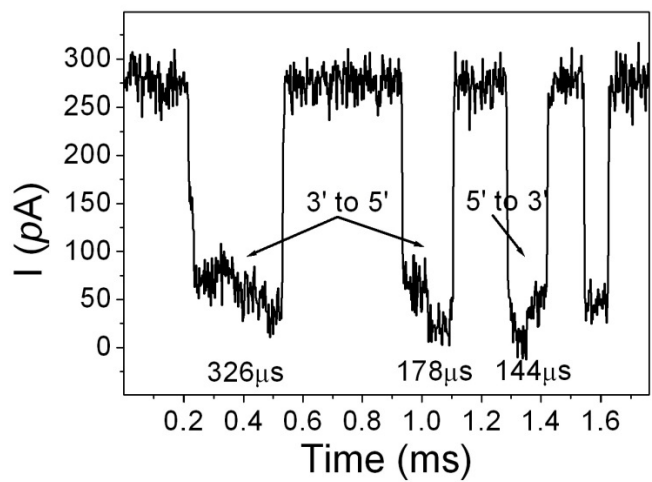

**Fig. 6**. Effect on current flow of ssDNA passage through an αHL channel. 5'$A_{50}C_{100}$3' ssDNA was introduced into the *cis* chamber (where the cap of the ion-channel is located). Buffer concentration = 1M KCl, applied voltage = +210 mV, amplifier bandwidth = 100 kHz. The channel's open state is interrupted by four brief closures (translocation events). The first and second events show two closure substates indicating the translocation of ssDNA from the 3' to the 5' end. The third event also shows two substates which, however, are reversed, indicating passage from the 5' to the 3' end. In the fourth event, the $A_{50}$ and $C_{100}$ sections of the molecule are indistinguishable

It has been shown by several research groups that, for a given polynucleotide, translocation events generally fall into two categories, with one group of events showing less current blockade than the other. This grouping has been suggested to arise from translocation of the molecule in either of two orientations, namely, 3' → 5' or 5' → 3', or it might represent the translocation of the polymer in one of two different structural conformations.[5-9] An asymmetric interaction between the polynucleotide and the internal pore structure has been suggested as the cause of the observed grouping in those occasions where the entry of the molecule into the pore from its 5' end produces a larger current blockade than when the 3' end enters the pore first.[12]

In our 5'$A_{50}C_{100}$3' ssDNA sample, keeping in mind that the C-nucleotide is smaller than the A-nucleotide, the 100-base-long C-segment is expected to drop the ionic current somewhat less than the 50 base-long A-segment does. The order in which the high and low portions of the bi-level signal occur during each translocation event depends on whether the 3'- or the 5'-end of the molecule enters the nano-pore first. Of the 29% bi-level events observed in the aforementioned experiment, nearly three-quarters were associated with the C-segment entering first, during which events the average normalized current, $<I_{blocked}/I_{open}>$, was 0.37 ± 0.09 for the C-segment and 0.17 ± 0.04 for the A-segment. In the remaining quarter of the bi-level events (associated with the A-segment entering first) the average normalized current was 0.20 ± 0.03 for the C- and 0.12 ± 0.04 for the A-segment. These results are consistent with the aforementioned suggestion that the entry of polynucleotides into the pore from the 5'-end causes a larger current blockage than entry from the 3'-end.

**7. Effect of increased voltage**. Both the capture rate of the DNA molecules and the translocation speed through a single ion-channel were found to be strong functions of the applied voltage $V$, as has been reported by other researchers as well.[5] The three frames in Fig. 7 represent the translocation of 120-base-long 5'(AC)$_{60}$3' DNA molecules under the influence of three different voltages. At $V = $ 150mV, the average open-channel current is $<I_{open}> \sim$160pA, and many translocation events are seen to occur. At $V=$ 120mV and $<I_{open}> \sim$ 130pA, the translocation rate has declined, and at $V=$ 90mV, $<I_{open}> \sim$ 90pA, the events are relatively rare.

The observed behavior may indicate that the increased voltage has somehow directed a large number of DNA molecules (that would otherwise drift away from the vicinity of the pore) toward the ion-channel. This, in turn, is a clue concerning the transport mechanism of DNA molecules toward the pore, namely, that the drift of the molecule is not entirely controlled by thermal diffusion, and that the chamber geometry and placement of the electrodes can influence (and perhaps even control) the motion of DNA strands toward the ion-channel. When considering this particular method of molecular readout in the context of data storage, it must be remembered that controlled molecular transport is of utmost significance, as individual macromolecules are required to travel between their parking spots and the read/write stations.

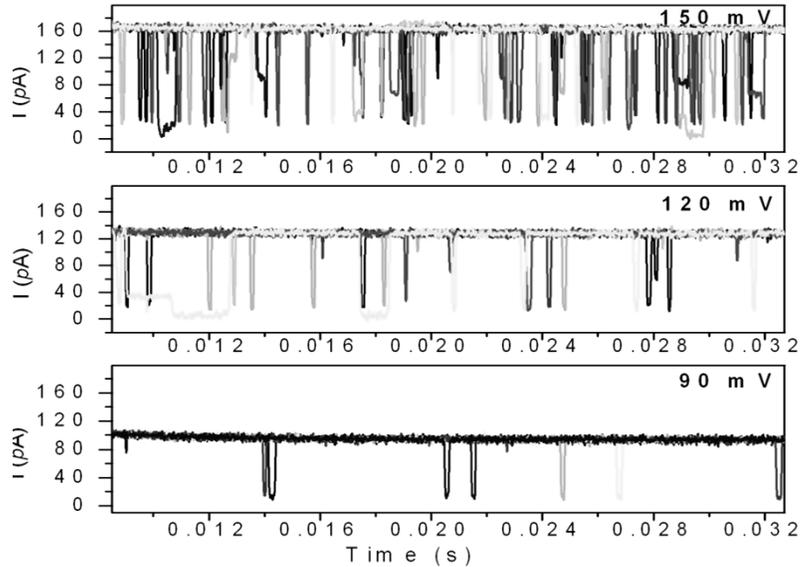

**Fig. 7**. Translocation of 120-base-long 5'(AC)$_{60}$3' DNA molecules under different applied voltages. At $V=$ 150mV, $<I_{open}> \sim$ 160pA and translocation events are frequent. At $V=$ 120mV, $<I_{open}> \sim$ 130pA and translocation rate has declined. At $V=$ 90mV, $<I_{open}> \sim$ 90pA and translocation events are rare.

Figure 8 shows some statistics of the translocation experiment depicted in Fig. 7. Figure 8(a) shows that at lower applied voltages the ion-channel is open (i.e., not clogged by a passing molecule) for a larger fraction of time than at higher voltages. According to Fig. 8(b), the translocation rate (i.e., number of events per second), whether complete (■) or incomplete (●), is an increasing function of the applied voltage. The total translocation rate (▲), which is the sum of complete and incomplete translocations per second, increases nearly three-fold in going from $V=$ 120mV to 150mV.

Figure 9 shows the distribution of the event duration $\Delta T$ versus the current blockage in the experiment of Fig. 7. These data indicate that, at higher applied voltages, there is less current blockage, perhaps because the molecules are likely to be linearized, thus presenting a smaller cross-section to the pore. Also, at higher applied voltages the translocation duration is reduced, meaning that the molecules travel faster through the ion-channel. The speed of the molecules passing through the nano-pore is seen to be roughly proportional to the applied voltage, as has been reported by other research groups.[5,9] Faster translocation, of course, is desirable in data storage applications as it results in a greater data-transfer rate, so long as the SNR remains reasonably high at the correspondingly larger bandwidth.

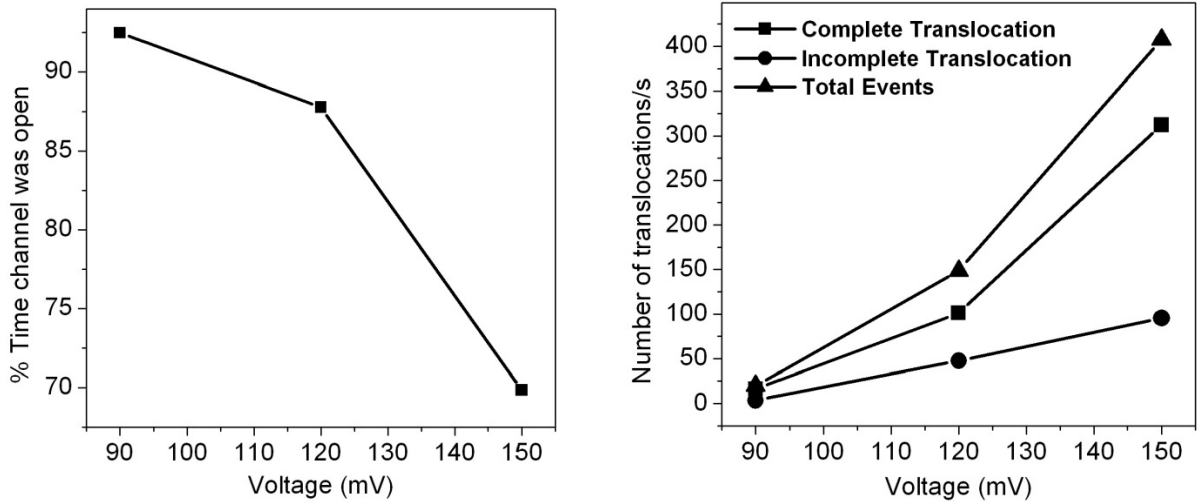

**Fig. 8**. Characteristics of translocation events in the experiment depicted in Fig. 7. (Left) Fraction of time during which the ion-channel is open (i.e., not clogged) as function of the applied voltage $V$. (Right) Translocation rate versus the applied voltage.

**Fig. 9**. Translocation statistics in the experiment depicted in Fig. 7. The percentage current blockage is shown on the horizontal axis, while the event duration appears on the vertical axis. The cluster of triangles (▲) represents the case of $V = 150$mV, circles (●) correspond to $V = 120$mV, and squares (■) represent the case of $V = 90$mV.

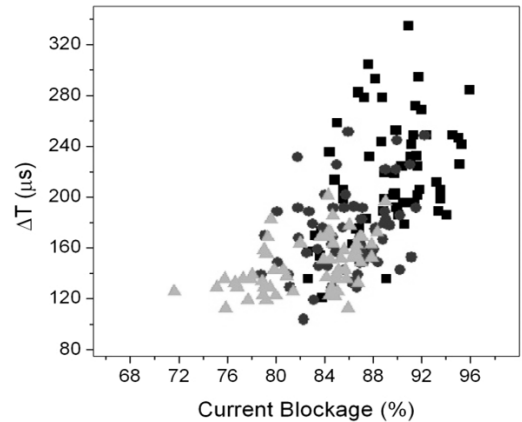

**8. Experiments with multiple nano-pores**. In a typical nano-pore experiment, αHL proteins are removed (by perfusion) from the *cis* chamber immediately after the formation of a single nano-pore. Given sufficient time, however, additional ion-channels reconstitute themselves in the same lipid membrane. In one experiment, the lipid bilayer had a diameter of 20μm, resulting in a pair wise separation of at most 20μm between nano-pores. Figure 10 shows a typical trace of the electrolytic current across the membrane where, initially, the total current is ~400 pA, indicating the presence of three open nano-pores. With all three channels open, the rate of capture and/or translocation is relatively high (31.8 events per second). When one of the pores is temporarily clogged, the current drops to ~290pA, and the rate of capture/translocation through the two remaining open pores drops to ~19.7 per second. Similarly, when two of the nano-pores are clogged, the current drops to ~190pA and the event rate declines to ~10.3 per second. This data indicates that, for each nano-pore, the electrolytic current is ~130pA in the open-channel state and ~30pA in the clogged state. The observed capture and/or translocation rate ratio of nearly 3 : 2 : 1 confirms that the three ion-channels operate independently of one another on drifting DNA molecules. We have observed the same behavior repeatedly: the event rate increased after reversing the applied voltage to clear the clogged channel(s), only to drop again with the next clogging.

**Fig. 10**. A typical trace of the electrolytic current obtained when three ion-channels (reconstituted in the same lipid membrane) operate simultaneously and independently of each other. When a DNA molecule is stuck in one of the nano-pores, the current drops to ~190pA. With two nano-pores similarly clogged, the total current across the membrane is ~190pA. The complete translocation events are found to have average rates of 12.6, 6.2, and 3.3 events per second, with zero, one, or two nano-pores clogged, respectively. The corresponding rate for partial events in these three cases are 19.2, 13.5, and 7.0 events/second, leading to total event rates of 31.8, 19.7 and 10.3 per second.

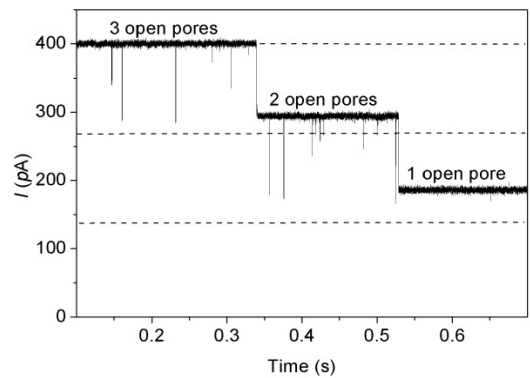

**9. Concluding remarks**. As a first step toward constructing a macromolecular data storage system, we have demonstrated the feasibility of conducting experiments with fairly short strands of DNA in a miniature read station. We showed that the A- and C-segments of a $5'A_{50}C_{100}3'$ ssDNA molecule can be distinguished during translocation through an αHL ion-channel. Needless to say, many hurdles must be overcome before this system can be implemented as an alternative to existing storage technologies.

The ultrahigh capacity is an obvious advantage of such molecular storage devices, but the data-rates require substantial improvement. If the $5'A_{50}C_{100}3'$ molecules, which represent 2-bit sequences of binary information, take ~150μs to pass through a nano-pore, the corresponding data-transfer rate of only 12 kbit/s would leave a lot to be desired. On the other hand, if the technology could improve to the point that individual bases could be detected during translocation (perhaps through a shorter, more robust, solid-state nano-pore),[13] then the achievable rate of ~1 Mbit/s would not be out of bounds. Moreover, if thousands of read stations could be made to operate in parallel within the same chip, the overall data-transfer rate could approach the respectable value of several Gbits/s.

Access to data is another issue that requires extensive research. Our preliminary calculations show that a 1 Mbyte-long macromolecule, enclosed in a 200 nm diameter spherical shell (e.g., a liposome), can move electrophoretically across a 1.0 $cm^2$ chip in ~1.0 millisecond under a 10 V potential difference. Whether such molecules could be written and packaged on demand, within an integrated micro-chip, at high speed, and on a large scale, are questions to which only future research can provide satisfactory answers. Stability of the molecules over extended periods of time should obviously be a matter of concern. The lipid membrane and the proteinaceous ion-channel, both being of organic origin, are ill-suited for practical data storage systems; they must eventually be replaced with robust, solid-state equivalents.[13] Despite all the difficulties, it is our hope that the proposed scheme will encourage debate in the pursuit of an alternative path to conventional approaches to data storage.

**Acknowledgements**. The authors are grateful to Professors Raphael Gruener, Michael Hogan, and Nasser Peyghambarian of the University of Arizona, and Joseph Perry and Seth Marder of the Georgia Institute of Technology, for many helpful discussions. This work has been supported by the *Office of Naval Research* MURI grant No. N00014-03-1-0793 and by the *National Science Foundation* STC Program, under Agreement No. DMR-0120967.

# Appendix

## Experimental conditions

The buffer used in our experiments was 1M KCl, 10mM HEPES (pH ~ 8.0), pH balanced by KOH, and the lipid was diphytanoyl phosphatidycholine (DPC). The Ag-AgCl electrodes were made by dipping a silver wire in Clorox. The patch-clamp amplifier was Axon Instrument's Axopatch 200B. Two different oligodeoxynucleotide samples were used in our experiments: the $5'A_{50}C_{100}3'$, with a total of 150 base-units per molecule, was used in the experiment described in Fig. 6. All other experiments (reported in Figs. 7-10) used $5'(AC)_{60}3'$, with a total of 120 bases per molecule. Both custom-synthesized, PAGE (Poly Acrylic Gel Electrophoresis) grade samples were purchased from *Midland Certified Reagent Company* (Midland, Texas). The samples were suspended in TE buffer (10mM Tris / 1mM EDTA) at pH 8.0, before being released into the cis chamber, where the final concentration of $5'A_{50}C_{100}3'$ was 7.15 nM/ml, while that of $5'(AC)_{60}3'$ was 14.3 nM/ml.

Prior to filling the chambers with buffer, 2μL of a lipid solution in hexane (1.5mg/mL) is released in the vicinity of the 20μm aperture between the *cis* and *trans* chambers. The chambers are then dried under a mild stream of nitrogen, thus allowing a thin layer of lipid to coat the surrounding walls without clogging the aperture. Care must be taken to completely evaporate the hexane, as even trace amounts in the aperture area can ruin the bilayer. We found the bilayer that must cover the aperture between the two chambers does not form properly when applied to a PDMS surface. Although PDMS is hydrophobic – a requirement for this application – its porosity ultimately destroys the lipid membrane a few minutes after the formation of a bilayer. In the read station depicted in Fig. 4, the needle-like end of the tube connecting *cis* and *trans* chambers had to be covered with a Teflon cap (aperture diameter ~ 20μm) to promote the proper formation of a lipid membrane.

The cis and trans chambers (as well as the tube that connects them) are subsequently filled with buffer. We soaked 1.5 mg of lipid in 5 μL of hexadecane in a test tube for 60-90 seconds; the left-over hexadecane was then completely drained, leaving a viscous lipid at the bottom of the tube. A micro-capillary tube (length ~ 1mm, inner diameter ~300μm) was filled with this viscous lipid in such a way as to form a convex meniscus at the tip of the capillary. Using a micro-manipulator, this lipid meniscus was brushed across the 20μm aperture with a slight pressure. A 60 Hz, 5 mV square wave was then applied between the Ag-AgCl electrodes across the bilayer (the so-called seal test) to verify that the bilayer is adequate. The above procedure succeeds in creating a proper bilayer across the aperture in about 90% of the trials.

Once a stable bilayer was obtained, 40 ng of α–haemolysin (αHL) protein was dissolved in 2μL of buffer and added to the cis chamber. Applying a potential ($V$ = 120 mV) across the bilayer enables one to observe the reconstitution of an ion-channel into the bilayer, an event which is indicated by an abrupt jump of the current from 0 to ~120pA. The observed voltage to current ratio of ~1.0 GΩ is consistent with the single-channel conductance of αHL nano-pores under similar conditions, as reported in the literature.[5] Self-assembly of a single ion-channel within the lipid membrane typically takes 20-30 minutes, after which the excess αHL is removed by perfusion of 10 mL of fresh buffer.